\documentclass[12pt]{article}
         
\usepackage{amsmath,amssymb} % amsmath
\usepackage{prepicte,pictex,postpict}     
\usepackage[mathscr]{eucal}       
\usepackage{latexsym,amsfonts,amstext,verbatim}            
\usepackage[dvips]{graphicx}

\newcommand{\ccomma}{\raisebox{.5ex}{,}\relax}

\newcommand{\bsm}{\boldsymbol}
\newcommand{\AB}{Aharonov-Bohm}
\newcommand{\rmi}{\mathrm{i}}
\newcommand{\rme}{\mathrm{e}}

\newcommand{\bR}{\mathbb{R}}
\newcommand{\bC}{\mathbb{C}}

\newcommand{\frD}{\mathfrak{D}}
\newcommand{\frH}{\mathfrak{H}}

\newcommand{\cW}{\mathscr{W}}

\begin{document}

 \title{\large\bf Can the \AB{} effect be used to detect or refute
Superseparability?}

\author{ {R N Sen}\\[2mm]
{\normalsize Department of Mathematics}\\
{\normalsize Ben-Gurion University, 84105 Beer Sheva, Israel}\\[2mm]
{\normalsize E-mail: rsen@cs.bgu.ac.il}}

\date{\normalsize 31 January 2010}

\maketitle
\begin{abstract}
\thispagestyle{empty}

In 1988, Reeh showed that the representation of the canonical
commutation relations that corresponds to the Aharonov-Bohm effect
depends on the magnetic flux $\Phi$. It can be integrated to a
representation of the Weyl group only if the flux is quantized.  It
follows from Reeh's analysis that representations for $\Phi_1\neq
\Phi_2$ are generally inequivalent. As a result, two identical
charged bosons may be found in inequi\-valent representations of the
CCR.  If unitary inequivalence is a restriction on superposability,
then these two particles should not feel each other's presence even
when they are in close physical proximity!  If they \emph{do} feel
each other's presence, then unitary inequivalence is \emph{not} a
restriction on superposability, and the question arises: what does
unitary inequivalence mean?  This paper suggests an experiment that
can distinguish between these two possibilities and provides a brief
account of the theory behind it, which depends upon the subtle notion
of self-adjointness of unbounded operators.

\end{abstract}\pagebreak

\hfill
\begin{minipage}[t]{7cm}
{\small\emph{\ldots you cannot occupy two places in space
simultaneously.  That is axiomatic.}

\vspace{2mm}\hfill{\small Hurree Babu, in Rudyard Kipling's \emph{Kim}}}
\end{minipage}

\section{Introduction}
\emph{Superseparability} may be defined as the polar opposite of
entanglement: two identical charged bosons, with state vectors that
have considerable spatio-temporal overlap, are unable to feel each
other's presence because the state vectors lie in disjoint Hilbert
spaces and cannot be superposed.  Mathematically, this possibility
appears to be contained in von Neumann's Hilbert space formulation of
quantum mechanics, but whether or not it is realized in nature can
only be ascertained by experiment.  This note describes the principle
of a possible experiment based on the magnetic \AB{} effect.

The mathematical phenomena that suggest the experiment are subtleties
hidden in the notion of self-adjointness for unbounded operators.
Experience shows that these subtleties can generally be disregarded
in practical applications of quantum mechanics; among physicists,
only the mathematically minded are likely to be familiar with them. For
this reason, a brief review of the basic definitions and results, due
mostly to von Neumann, is provided in Section~\ref{SEC-SELF-ADJ}.  It
follows an essential review of the historical background in
Section~\ref{SEC-HIST}.  The paper begins with a description of the
experimental scheme in Section~\ref{SEC-SCHEME}; the theory of the
experiment is given, after the historical and mathematical
excursions, in Section~\ref{SEC-REEH}. If the experiment turns out to
be feasible, its results -- be they positive or negative -- will be
consequential, and implications of the possible results are discussed
briefly, in a non-speculative manner, in Section~\ref{SEC-DISCUSS}. 

\section{Scheme of the experiment}\label{SEC-SCHEME}

The purpose of the experiment is to determine whether or not two
beams of identical bosons, prepared in inequivalent representations
of the CCR, can interfere with each other.  A possible scheme for
such an experiment is shown in figure~1.

A coherent beam of charged bosons (e.g., $\alpha$-particles or
deuterons) from a source $S$ is split into two by a beam-splitter
$P$. One beam goes through the chamber $A$, the other through the
chamber $B$. Neither chamber contains a magnetic field, and the two
are electromagnetically isolated from each other. They contain the
magnetic flux lines $\Phi_{A,B}$ respectively (perpendicular to the
plane of the paper); at least one of these fluxes is continuously
variable over a certain range. The experiment consists of observing
changes in the interference pattern at the detector $D$ as
$\Delta\Phi=\Phi_B-\Phi_A$ is varied. 

\begin{figure}[ht]
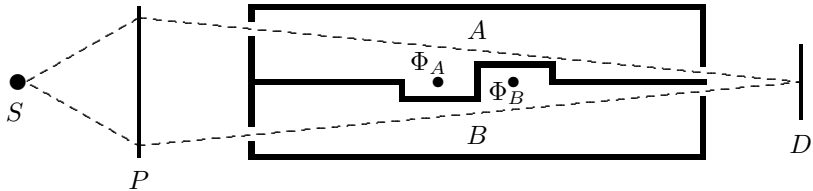


\vspace{5mm}

\beginpicture

\setcoordinatesystem units <1mm,1mm>%%

\setplotarea x from -75 to 45, y from -20 to 15

\linethickness=2pt

\putrule from -30 -10 to 30 -10
\putrule from -30 10 to 30 10
\putrule from -30 0 to -10 0
\putrule from 10 0 to 30.4 0
\putrule from -10 0.3 to -10 -2.3  % vert OK
\putrule from -10.4 -2.3 to 0.4 -2.3
\putrule from 0 -2.3 to 0 2.3        % vert OK
\putrule from -0.4 2.3 to 10.4 2.3
\putrule from 10 2.3 to 10 -0.3       % vert OK
\putrule from -30 -10.3 to -30 -8  % vert OK
\putrule from -30 -6 to -30 6    % vert )OK
\putrule from -30 8 to -30 10.3    % vert OK
\putrule from 30 -10.3 to 30 -2    % vert
\putrule from 30 2 to 30 10.3      % vert 

\put {\circle*{4}} [B1] at -4.5 0
\put {\circle*{4}} [B1] at 5.5 0
\put {\circle*{6}} [B1] at -60 0

\linethickness=1pt

\putrule from -45 -10 to -45 10
\putrule from 43 -5 to 43 5

\linethickness=0.5pt

\setdashes <1mm>

\plot -45 8.5  43 0  -45 -8.5 /
\plot -45 8.5  -60 0  -45 -8.5 /
\footnotesize
\put {$S$} at -61.5 -4
\put {$P$} at -45 -13
\put {$D$} at 43 -8.3
\put {$\Phi_A$} at -6.5 2.5
\put {$\Phi_B$} at 4 -1.5
\put {$B$} at 0 -7
\put {$A$} at 0 7

\endpicture
\vspace{5mm}
\caption{Scheme for a noninterferometer}
\end{figure}

As will be shown in Section~\ref{SEC-REEH}, von Neumann's Hilbert
space formulation of quantum mechanics \cite{vN1932} suggests
that the two beams should interfere \emph{only} when
$\Delta\alpha=q\Delta\Phi/2\pi$ is an integer, where $q$ is the
charge of the boson, which will be $-e$ for deuterons and and $-2e$
for $\alpha$-particles. (We use units in which $\hbar=c=1.)$ For
non-integral values of $\Delta\alpha$, the two beams will belong to
disjoint Hilbert spaces and should not -- if the phrase
\emph{disjoint Hilbert spaces} has physical meaning -- be able to
interfere with each other.  For this reason, the scheme of the figure
is called a \emph{noninterferometer}.\footnote{The terms
\emph{superseparability \emph{and} noninterferometer} were introduced
in \cite{SEN2010}.} It is also possible that the above interpretation of
quantum mechanics is invalid, and that both beams belong to the same
Hilbert space. In that case the interference pattern should merely
shift, as $\Delta\alpha$ is varied, returning to the original state
when $\Delta\alpha$ has changed exactly by unity; the fringe shift
should be periodic, with period $1$. If the experiment is realizable,
the case $\Phi_A=\Phi_B\neq 0$ will correspond to the standard
Aharonov-Bohm effect, so that this effect may be used to test the
electromagnetic isolation of the chambers $A$ and $B$. The reader is
referred to the monograph by Peshkin and Tonomura \cite{P-T1989} for a
historical account of the Aharonov-Bohm effect, and to the review
article \cite{T2005} and monograph \cite{T1999} for the
decisive experiments.

It should be recalled that if a flux $\Phi$ is quantized, then
$2e\Phi = 2\pi n$, where $n$ is an integer, so that $\Delta\alpha =
(n_B-n_A)/2$ for deuterons and $\Delta\alpha = n_B-n_A$ for
$\alpha$-particles. Therefore superseparability will never be
observed with $\alpha$-particles if both fluxes are quantized. 

We shall end this section with a few reservations and a remark. The
configuration described above is an ideal which may be hard to
realize in the laboratory; that is why we have called it the `scheme'
of an experiment. It is unlikely that the chambers $A$ and $B$ can be
perfectly isolated from each other.  It may not be possible to
prepare a state with a sharp value of $\alpha$ if the flux is not
quantized. Finally, the theory of Section~\ref{SEC-REEH} would be
applicable only if the interiors of chambers $A$ and $B$ are, for
purposes of the experiment, reasonable approximations to the punctured
plane. On the other hand, we know -- if only by hindsight -- that the
\AB{} effect can be observed under conditions that are less than ideal.
Therefore the possibility that superseparability may also be testable
under less than ideal conditions should not be ruled out of hand.

\section{Historical background}\label{SEC-HIST}

We begin by recalling two basic facts. (i) The Born-Jordan
commutation relation $[p,q] = -\rmi\,\!I$ cannot be represented by
finite-dimensional matrices if $I$ is required to be the identity
matrix. (ii) If it is represented on an infinite-dimensional Hilbert
space $\frH$ with $I$ as the identity operator, then at least one of
$p$ and $q$ must be represented by an unbounded operator (see, for
example, \cite{SEN2010}.\footnote{All our Hilbert spaces will be over
the complex numbers, and will have countable orthonormal bases.} 

Unbounded operators are not defined everywhere on a Hilbert space,
and are discontinuous wherever they are defined. They give rise to
mathematical phenomena that are not encountered in the theory of
finite dimensional matrices, and it requires considerable effort to
invest with meaning even the simplest of assertions, such as
$[A,B]=0$, if $A$ and $B$ are unbounded. The basic structures of
quantum mechanics, namely matrix mechanics, wave mechanics and
transformation theory were laid down in 1925--27,\footnote{Dirac's
\emph{Principles of Quantum Mechanics} \cite{D1930} and Heisenberg's
\emph{Physical Principles of Quantum Mechanics} \cite{H1930}
were both published in 1930.} but unbounded operators began to be
explored only in 1929--1930 \cite{vN1929-1930,S1932}. The
`first quantum revolution' (this term is due to Aspect \cite{A2004}) was
completed while unbounded operators were still terra incognita even
to mathematicians. In retrospect, one is struck by the fact that
transformation theory could be developed with scant understanding of
the operators that were to be transformed. By what magic was this
achieved? 

The answer lies in an ansatz due to Hermann Weyl and a theo\-rem
proven by von Neumann.
 
In 1928, Weyl published his book \emph{Gruppentheorie und
Quantenmechanik} \cite{W1928}. In this book he replaced the canonical
commutation relations (CCR) for $N$ degrees of freedom by a
$2N$-parameter Lie group, which had the CCR as its Lie algebra.  This
group has become known as the \emph{Weyl group}, and we shall denote
it by $\cW_N$. We shall give the argument for $N=1$; the general case
merely requires a cumbersome modification of the notation (see \cite{W1928},  
pp.\ 272--276).

Let $a, b \in \bR$ and define, formally,  
\begin{equation}\label{WEYL-GROUP-1}
u(a) = \exp\,(\rmi ap),\qquad v(b) = \exp\,(\rmi bq).
\end{equation}
From the properties of the exponential function, it follows
that
\begin{equation}\label{WEYL-GROUP-2}
u(a)u(a^{\prime}) = u(a+a^{\prime}), \qquad
v(b)v(b^{\prime}) = v(b+b^{\prime}).
\end{equation}
Write $u(-a)= u(a)^{-1}$, $v(-b)=v(b)^{-1}$ and
$u(0)=v(0)=\mathbf{1}$. Formal computation yields the result
\begin{equation}\label{WEYL-GROUP-3} 
u(a)v(b)u(a)^{-1}v(b)^{-1} =
\rme^{\rmi ab}\mathbf{1}.  
\end{equation} 
By definition, the Weyl group $\cW_1$ consists of the set of elements
(\ref{WEYL-GROUP-1}), with multiplication defined by
(\ref{WEYL-GROUP-2}) and (\ref{WEYL-GROUP-3}). The element
$\mathbf{1}$ is the identity of the group. The group $\cW_1$ is
nonabelian and noncompact, with $\bR^2$ as the group manifold, and is
a Lie group. The same is true of the Weyl group $\cW_N$ for $N$
degrees of freedom, except that its group manifold is $\bR^{2N}$.

Being noncompact, the Weyl groups have no finite dimensional unitary
representations. In a unitary representation, the elements $u(a)$ and
$v(b)$ of $\cW_1$ are represented by unitary operators $U(a)$ and
$V(b)$ on the Hilbert space $\frH$, and similar statements hold for
$\cW_N$.\footnote{The definition of an infinite-dimensional unitary
representation includes a continuity condition that we have not
specified. The same condition is used in the definition of
one-parameter groups of unitaries.} A result known as Stone's theorem
asserts that a one-parameter group of unitaries $\{U(t)\}$ on a
Hilbert space has an infinitesimal generator $H$: $U(t) = \exp\,(\rmi
Ht)$, where $H$ is self-adjoint. It is bounded if $\{U(t)\}$ is
compact ($t\in S_1$, the circle) and unbounded if $\{U(t)\}$ is not
compact ($t \in \bR$).  A representation of $\cW_N$ defines,
uniquely, a representation of its Lie algebra -- the CCR -- by
self-adjoint operators.  In the representation so defined, at least
one member of any canonical pair $p, q$ is represented by an
unbounded operator.\footnote{Self-adjoint operators on
infinite-dimensional Hilbert spaces will be defined precisely in
{Section}~\ref{SEC-SELF-ADJ}. The exponential $\exp\,(\rmi At),\,
t\in\bR$ of the unbounded self-adjoint operator $A$ needs definition,
but we shall content ourselves with the statement that it turns out
to have the expected properties.}

In 1930 von Neumann proved that, for finite $N$, the Weyl group
$\cW_N$ has only one irreducible unitary representation
\cite{vN1930}. He gave the name \emph{Schr\"odinger operators} to the
representatives of the canonical variables $p_j, q_j$,
$j=1,\ldots,N$, and titled his paper `Die Eindeutigkeit
Schr\"odingersche Operatoren'. His result has become known as `von
Neumann's uniqueness theorem'.  If the CCR were equivalent to the
Weyl group, it would explain why quantum mechanics could be developed
ahead of the theory of unbounded operators without falling into gross
error.

A Lie group defines a unique Lie algebra, but the converse is not
true. The simplest examples are the covering groups of compact
non-simply-connected Lie groups. Examples of this phenomenon that are
relevant to elementary particle physics were unearthed by Michel 
as early as 1962 \cite{M1964}. The canonical commutation relations
are \emph{not} abstractly equivalent to the Weyl group; as we shall
see below, the $p_j,q_k$ will not even generate a Lie group unless
they are represented by self-adjoint operators.  However, the
requirement of self-adjointness cannot be met in some simple and
realizable physical situations.

\section{Self-adjointness}\label{SEC-SELF-ADJ}

Let $\frH$ be a Hilbert space and $A$ an operator on it. If there
exists a positive number $K$ such that $||A\psi|| \leq K ||\psi||$
for all $\psi\in\frH$, then $A$ is said to be \emph{bounded}. If no
such $K$ exists, then $A$ is said to be \emph{unbounded}. An
unbounded operator $A$ is not defined everywhere on $\frH$; the
subset $\frD(A)\subsetneq\frH$ on which it is defined is called the
\emph{domain} of $A$. If $\frD(A)$ is not dense in $\frH$ then $A$ is
not (yet) mathematically manageable, and one generally assumes that
$A$ is \emph{densely defined}, i.e., $\frD(A)$ is dense in $\frH$.

In the rest of this section we shall deal only with unbounded
operators, and therefore the adjective `unbounded' will be omitted.

An operator $A$ is called \emph{closed} if the set of ordered pairs
$\{(\psi, A\psi)|\psi\in\frD(A)\}$ is a closed subset of
$\frH\times\frH$. An operator $A_1$ is an \emph{extension} of $A$ if
$\frD(A)\subset\frD(A_1)$ and $A_1\psi=A\psi$ for $\psi\in\frD(A)$;
one writes $A \subset A_1$. An operator is called \emph{closable} if
it has a closed extension. Every closable operator $A$ has a smallest
closed extension, which is denoted by $\bar{A}$.

In matrix theory, the adjoint is defined by $(T\bsm{x}, \bsm{y}) =
(\bsm{x}, T^{\star}\bsm{y})$. In operator theory, one has to take
domains into consideration. Let $\varphi,\xi\in\frH$ such that
$(A\psi,\varphi)=(\psi,\xi)$ for all $\psi\in\frD(A)$, and define
$A^{\star}$ by $A^{\star}\varphi=\xi$. Then $\frD(A^{\star})$ is
precisely the set of these $\varphi$. One can show that if $A$ is
densely defined, then $A^{\star}$ is closed. Furthermore, $A^{\star}$
is densely defined if and only if $A$ is closable, and if it is, then
$(\bar{A})^{\star} = A^{\star}$.

If $\mathfrak{D}(A)\subset\mathfrak{D}(A^*)$ and $A\varphi =
A^*\varphi$ for all $\varphi\in\mathfrak{D}(A)$, then $A$ is called
\emph{symmetric}.\footnote{Von Neumann used the term
\emph{Hermitian}, but current usage seems to limit this term to
operators on finite-dimensional vector spaces.} If $\mathfrak{D}(A) =
\mathfrak{D}(A^*)$ and $A\varphi=A^*\varphi$ for all $\varphi \in
\mathfrak{D}(A)$, then $A$ is called
\emph{self-adjoint}.\footnote{Von Neumann used the term
\emph{Hermitian hypermaximal}.} Self-adjoint operators form a
subclass of symmetric operators.

 A symmetric operator may have no self-adjoint extension, it may have
many self-adjoint extensions, or it may have only one. In the last
case, it is called \emph{essentially self-adjoint}. One can show that
if $A$ is essentially self-adjoint, then its closure $\bar{A}$ is
self-adjoint, i.e., $\bar{A}$ is the unique self-adjoint extension of
$A$.

The fundamental differences between symmetric and self-adjoint
operators are:

\begin{enumerate}

\item The spectrum of a self-adjoint operator is a subset of the real
line, whereas the spectrum of a symmetric operator is a subset of the
complex plane; a symmetric operator is self-adjoint if and only if
its spectrum is a subset of the real line.

\item A self-adjoint operator can be exponentiated, i.e., if $A$ is
self-adjoint then $\exp\,(\mathrm{i} tA)$ is defined for all $t \in
\mathbb{R}$; a symmetric operator which is not self-adjoint
\emph{cannot} be exponentiated.

\end{enumerate}

The representation problem for the CCR (one degree of freedom) may
now be formulated as follows: \emph{Find all pairs of essentially
self-adjoint operators $P, Q$, densely defined on a common domain
$\frD$, such that $[P,Q]\varphi = -\rmi\varphi$ for every
$\varphi\in\frD$}. There are infinitely many such (inequivalent)
pairs; the interested reader is referred to Schmudgen \cite{SCH1983}
for details, and for references to earlier works.

If $A$ and $B$ are self-adjoint, are defined on a common dense domain
$\mathfrak{D}$ and commute on $\mathfrak{D}$, then $\exp\,
(\mathrm{i}aA)$ and $\exp\,(\mathrm{i}bB)$ are defined for all $a,b
\in \mathbb{R}$ and commute. However, if $A$ and $B$ are merely
essentially self-adjoint, are defined on $\mathfrak{D}$ and commute
on $\mathfrak{D}$, then $\exp\,(\mathrm{i}a\bar{A})$ and $\exp\,
(\mathrm{i}b\bar{B})$ \emph{do not necessarily commute}. This fact,
which at first seems highly counterintuitive, was unearthed by Nelson
in 1958; for details and references, see Reed and Simon
\cite{R-S1972,R-S1975}.

We shall conclude this section with an example. The group of
isometries of $\bR^2$ consists of translations and rotations. The
group of isometries of the punctured plane $\bR^2\setminus\{O\}$ is
the group of rotations about the origin $O$. What happens to the
translation operators on $\bR^2$, namely $\exp\,(\rmi ap_x)$ and
$\exp\,(\rmi bp_y)$, $a,b \in \bR$ (where
$p_x=-\rmi\partial/{\partial x},\;p_y=-\rmi\partial/{\partial y}$),
when the origin is excised?

The operators $\partial/{\partial x},\,\partial/{\partial y}$ are
defined on sets of differentiable functions. A function which is
differentiable on $\bR^2$ is necessarily differentiable on
$\bR^2\setminus\{O\}$, but the latter has a richer supply of
differentiable functions than $\bR^2$, e.g., the function
$r^{-1}\exp\,(-r^2/2)$ (which is also square-integrable). Restricting
the domain enlarges the set of differentiable functions on which
$p_x$ and $p_y$ are defined. This enlargement changes the
\emph{spectra} of these operators, which in turn leads to the failure
of self-adjointness and exponentiability.

\section{Theory of the experiment}\label{SEC-REEH}

In 1988, Helmut Reeh showed that that the `Nelson phenomenon' could
be found in the Aharonov-Bohm effect \cite{R1988}.  For brevity, let
us call a spinless particle of charge $q$ moving in a plane
perpendicular to a trapped magnetic flux -- the classical
Aharonov-Bohm example -- an \emph{AB-particle}. Owing to cylindrical
symmetry, the motion of an AB-particle is essentially
two-dimensional.  Its canonical operators may be written, formally,
as 

\begin{equation}\label{AB-OP-1} 
\bsm{p} = -\mathrm{i}\frac{\partial}{\partial\bsm{x}} +
q\bsm{A},\;\; \bsm{q} = \text{multiplication by}\;\bsm{x}.  
\end{equation} 
\noindent
Boldface symbols denote 2-vectors in the $XY$-plane. The vector
potential $\bsm{A}$ (up to a gauge) can be written, in terms of the
magnetic flux $\Phi$, as 
\begin{equation}\label{VP-FLUX} 
\bsm{A} = \frac{\Phi}{2\pi r}\bsm{e}, 
\end{equation} 
where $r = (x^2+y^2)^{1/2}$ and $\bsm{e}$ is the unit vector at
$(x,y)$ tangent to the circle $r = \text{const}$: 
\begin{equation*} 
\bsm{e} = \left(-\frac{y}{r}\ccomma{}\; \frac{x}{r}\right)\cdot 
\end{equation*} 
We shall set $ \alpha = q{\Phi}/{2\pi}$ and use (\ref{VP-FLUX}) to
rewrite the quantities $\bsm{p}$ of (\ref{AB-OP-1}) as 
\begin{equation}\label{P-A} 
\bsm{p}^{\alpha} = -\rmi\frac{\partial}{\partial\bsm{x}} +
\alpha\bsm{e}, 
\end{equation} 
where the $\alpha$-dependence of $\bsm{p}$ has been rendered explicit
on the left. The problem is to define the formal quantities
$p_x^{\alpha}$ and $p_y^{\alpha}$ in (\ref{P-A}) as operators on the
Hilbert space $L^2(\mathbb{R}^2 \setminus O) = L^2(\mathbb{R}^2)$;
excision of a single point, here the origin $O$, has no real effect
on an $L^2$-space, but -- as we have seen earlier -- changing the
domains of differentiation operators ever so slightly can have
drastic consequences.  Reeh chose, for the common domain of
$p_x^{\alpha},\, p_y^{\alpha}$, the space $\mathscr{D}(\mathbb{R}^2
\setminus O)$ of smooth functions with compact support on
$\mathbb{R}^2\setminus O$. The space $\mathscr{D}(\mathbb{R}^2
\setminus O)$ is dense in $L^2(\bR^2)$, and $p_x^{\alpha}$ and
$p_y^{\alpha}$ are distributions on it.  If $\varphi \in
\mathscr{D}(\mathbb{R}^2 \setminus O)$, then it follows from
$\text{curl}\,\bsm{A} = 0$ that $[p_x^{\alpha},\,
p_y^{\alpha}]\varphi = 0$. 

Consider now the equation
\begin{equation}\label{EIGEN}
{p}^{\alpha}_x\varphi = \left(-\rmi\frac{\partial}{\partial x} -
\alpha\frac{y}{x^2+y^2}\right)\varphi
=  \lambda\varphi.
\end{equation}
It is a linear homogeneous differential equation of the first order
which can be solved explicitly for any $\lambda\in\bC$, and the same
holds for the equation ${p}_y^{\alpha}\psi = \lambda\psi$.  The
solutions do not have compact support. By exploiting these solutions,
Reeh established the following results \cite{R1988}:

\begin{enumerate}

\item The operators $p_x^{\alpha}$ and $p_y^{\alpha}$ 
are not self-adjoint; they are essentially self-adjoint.

 \item Let $\bar{p}_x^{\alpha}$ and $\bar{p}_y^{\alpha}$ be their
self-adjoint extensions, and define 
\begin{equation*}
V_x^{\alpha}(a) = \exp\,(\rmi a\bar{p}_x^{\alpha}),
\qquad
V_y^{\alpha}(b) = \exp\,(\rmi b\bar{p}_y^{\alpha}).
\end{equation*} 
Then 
\begin{equation}\label{REEH-COMM}
V_x^{\alpha}(a)V_y^{\alpha}(b)V_x^{\alpha}(a)^{-1}
V_y^{\alpha}(b)^{-1} 
 = \rme^{\rmi(\pi\alpha/2)\cdot[\epsilon(x) -
\epsilon(x+a)][\epsilon(y) - \epsilon(y-b)]}\,I, 
\end{equation}
where $I$ is the identity operator, and
\begin{equation*}
\epsilon(t) = \left\{\begin{array}{rl}
               1&\;\; t>1\\[2mm]
               -1&\;\;t<1.
              \end{array}
              \right.
\end{equation*}
\end{enumerate}
Note that the product $[\ldots][\ldots]$ in the exponent on the
right-hand side of (\ref{REEH-COMM}) can only assume the values
$0,\pm4$, so that the entire right-hand side can only assume the
values $I, \exp\,(\pm 2\pi\rmi\alpha)I$. It follows that if $\alpha$
is an integer, then the right-hand side of (\ref{REEH-COMM}) equals
the identity operator $I$ \emph{for all admissible} $x,y,a,b$, but
\emph{not} if $\alpha$ is not an integer; in this case the group
generated by the operators $\{x, y, \bar{p}_x^{\alpha},
\bar{p}_y^{\alpha}\}$ is no longer isomorphic with the Weyl group
$W_2$. Clearly, the groups generated by these operators for $\alpha =
\alpha_1, \alpha_2$ are not isomorphic with each other if
$\alpha_1-\alpha_2$ is not an integer, and therefore the
representations of the CCR (for two degrees of freedom) they define
are not unitarily equivalent. 

The experiment suggested in {Section}~\ref{SEC-SCHEME} is designed to
determine whether this mathematical inequivalence has observable
physical consequences.

\section{Interpretation of possible results}\label{SEC-DISCUSS}

\begin{enumerate}

 \item If superseparability is observed with charged boson beams,
then -- irrespective of the psychological effect of the observation
-- it will confirm that the notion of inequivalent irreducible
representations of the CCR (for a finite number of degrees of
freedom) is physically meaningful; vectors from two inequivalent
representations cannot be superposed upon each other. It should then
prompt the investigation of other possible effects that arise from
the existence of inequivalent irreducible representations of the CCR.
Note that the discussion is at the level of the first quantization.

 \item If, however, the phenomenon is \emph{not observed}, we shall
have to conclude that something basic is lacking in our understanding
of the linear space that underlies quantum mechanics: the question
that David Hilbert asked Rolf Nevanlinna in the late 1920's --
\emph{Tell me, Rolf, what is this Hilbert Space that the young people
are talking about?} -- may not yet have been answered to the physicist's
satisfaction.

 \item The theoretical considerations of Section \ref{SEC-REEH} do
not apply to fermions.  The creation-annihilation operators for a
fermion are bounded, and the canonical anticommutation
relations for a finite number of degrees of freedom have only one
irreducible unitary reprsentation. This was proved by Jordan and
Wigner in their very first paper on anticommutation relations in 1928.
Therefore one should not expect to find the phenomenon of
superseparability among fermions. If the experiment is performed with
electrons and superseparabi\-lity \emph{is} observed, it will pose
new and unsettling problems for theoretical physics. 

\end{enumerate}

\section*{Acknowledgements}

I would like to thank T Eisenberg, N Panchapakesan, H Reeh, H Roos, G
L Sewell and A Shimony for their comments on earlier versions of this
paper.

\end{document}